\documentclass[twocolumn,pra,nobalancelastpage,superscriptaddress,floatfix,nofootinbib]{revtex4}

\usepackage{amsmath}
\usepackage{graphicx}
\usepackage{amssymb}

\begin{document}




\newcommand{\non}{\nonumber}
\newcommand{\ssize}{\scriptsize}


\newcommand{\ie}{i.e.\ }
\newcommand{\etal}{{\em et al.}}


\newcommand{\tr}{\ensuremath{\mathrm{tr}}}
\renewcommand{\Pr}{\ensuremath{\mathrm{Pr}}}
\newcommand{\Sch}{\ensuremath{\mathrm{Sch}}}


\renewcommand{\>}{\ensuremath{\rangle}}
\newcommand{\<}{\ensuremath{\langle}}
\newcommand{\ket}{\ensuremath{\rangle}}
\newcommand{\bra}{\ensuremath{\langle}}
\renewcommand{\l}{\left}
\renewcommand{\r}{\right}
\renewcommand{\[}{\left[}
\renewcommand{\]}{\right]}
\renewcommand{\(}{\left(}
\renewcommand{\)}{\right)}


\newcommand{\eq}{\equiv}
\newcommand{\tp}{\ensuremath{\otimes}}


\newcommand{\A}{\ensuremath{{\cal A}}}
\newcommand{\B}{\ensuremath{{\cal B}}}
\newcommand{\C}{\ensuremath{{\cal C}}}
\newcommand{\D}{\ensuremath{{\cal D}}}
\newcommand{\E}{\ensuremath{{\cal E}}}
\newcommand{\F}{\ensuremath{{\cal F}}}
\newcommand{\G}{\ensuremath{{\cal G}}}
\newcommand{\I}{\ensuremath{{\cal I}}}
\newcommand{\LU}{\ensuremath{\mathbb{LU}}}
\newcommand{\R}{\ensuremath{{\cal R}}}
\renewcommand{\S}{\ensuremath{{\cal S}}}
\newcommand{\U}{\ensuremath{\mathbb{U}}}


\newcommand{\Kdel}{\ensuremath{K_{\Delta E}}}
\newcommand{\Khar}{\ensuremath{K_\mathrm{Har}}}
\newcommand{\Ke}{\ensuremath{K_E}}
\newcommand{\Ksch}{\ensuremath{K_\mathrm{Sch}}}
\newcommand{\Kd}{\ensuremath{K_D}}
\newcommand{\Khs}{\ensuremath{K_\mathrm{HS}}}
\newcommand{\Kcom}{\ensuremath{K_\mathrm{com}}}
\newcommand{\Kmax}{\ensuremath{K_{\max}}}
\newcommand{\Kt}{\ensuremath{K_\mathrm{T}}}


\newcommand{\cnot}{\textsc{cnot}}
\newcommand{\csign}{\textsc{csign}}
\newcommand{\swap}{\textsc{swap}}


\newtheorem{theorem}{Theorem}
\newtheorem{lemma}[theorem]{Lemma}
\newtheorem{proposition}[theorem]{Proposition}
\newcommand{\proof}{\noindent{\bf Proof:} }
\newcommand{\qed}{\hfill $\blacksquare$}
\newtheorem{axiom}{Axiom}
\newtheorem{property}{Property}
\newtheorem{conjecture}{Conjecture}
\newtheorem{definition}{Definition}

\title{Experimental requirements for Grover's algorithm in optical
quantum computation}

\author{Jennifer L.~Dodd}
\email{www.physics.uq.edu.au/people/jdodd}
\affiliation{Centre for Quantum Computer Technology, The University of
Queensland, QLD 4072, Australia}
\affiliation{School of Physical Sciences, The University of
Queensland, QLD 4072, Australia}
\affiliation{Institute for Quantum Information, California Institute
of Technology, Pasadena CA 91125, USA}
\author{Timothy~C.~Ralph}
\author{G.~J.~Milburn}
\affiliation{Centre for Quantum Computer Technology, The University of
Queensland, QLD 4072, Australia}
\affiliation{School of Physical Sciences, The University of
Queensland, QLD 4072, Australia}
\date{\today}

\begin{abstract}
The field of linear optical quantum computation (LOQC) will soon need
a repertoire of experimental milestones.  We make progress in this
direction by describing several experiments based on Grover's
algorithm.  These experiments range from a relatively simple
implementation using only a single non-scalable $\cnot$ gate to the
most complex, requiring two concatenated scalable $\cnot$ gates, and
thus form a useful set of early milestones for LOQC.  We also give a
complete description of basic LOQC using polarization-encoded qubits,
making use of many simplifications to the original scheme of Knill,
Laflamme, and Milburn~\cite{Knill01b}.
\end{abstract}
\maketitle

%
\section{Introduction}

%
In the next few years, we can expect to see demonstrations of basic
quantum gates in several implementations of quantum computation.  With
this in sight, it is natural to look ahead to what interesting quantum
circuits can be built out of a small number of one- and two-qubit
gates acting on a few qubits, as these circuits will provide
milestones on the way to full-scale quantum computation~\cite{QIST02}.

%
Grover's search algorithm~\cite{Grover96,Grover97} is a good candidate
for such a milestone.  It is a quantum algorithm identifying one of
$N$ elements, marked by an oracle, with order $\sqrt N$ uses of the
oracle.  When the search space consists of 4 elements, the algorithm
is guaranteed to produce the marked element after one use of the
oracle, compared to the 2.25 uses expected in a classical search.  We
will see that it can be implemented using only 7 one-qubit gates and 2
two-qubit gates, which makes it an excellent target once one- and
two-qubit gates have been mastered.  Not surprisingly, it was one of
the first algorithms to be experimentally implemented in nuclear
magnetic resonance quantum computing (Chuang, Gershenfeld, and
Kubinec~\cite{Chuang98} and Jones, Mosca, and Hansen~\cite{Jones98b}).

%
A promising quantum computing technology is the \emph{linear optical
quantum computation} (LOQC) scheme of Knill, Laflamme, and
Milburn~\cite{Knill01b} (see Gottesman, Kitaev, and
Preskill~\cite{Gottesman00} for an alternative approach).  In this
scheme, one-qubit gates are relatively straightforward.  While
implementing a scalable universal two-qubit gate such as a $\cnot$
remains a challenge, such a gate is likely to be demonstrated in the
next couple of years.  Already, a non-scalable $\cnot$ has been
approximately implemented by Pittman \etal~\cite{Pittman03}.  For
these reasons, it is important to establish some specific LOQC
milestones on the path toward building a large quantum computer, in
the form of some simple algorithms on a few qubits.

%
This pursuit is dogged by conceptual difficulties associated with
quantum algorithms on a very small number of qubits, summed up in the
question: What is the criterion for ``quantumness''?  A reasonable
criterion, particularly in the context of Grover's algorithm, is to
require a ``speed-up'' over the best classical algorithm.  However,
this notion can be hard to make sense of when the number of steps is
on the order of ten, rather than millions, and the problem can easily
be done by hand (not to mention by a GHz classical processor).
Furthermore, sometimes the reduction in the number of steps can be
achieved in an implementation whose physical requirements grow
exponentially with the number of qubits, trading off time for space.
The question of whether or not this counts as ``quantum'' has received
much attention (see, for example, Kwiat \etal~\cite{Kwiat00},
Bhattacharya, van den Heuvell, and Spreeuw~\cite{Bhattacharya02}).

Perhaps the best solution to this problem is a pragmatic one.  In the
quest to build a quantum computer large enough to provide a genuine
advantage over classical computers, two things must be achieved.
First, a fine level of quantum control must be demonstrated for both
single qubits and pairs of qubits.  Second, it will be necessary to
show that the number of components (qubits and gates) in a circuit can
be increased without insurmountable increases in difficulty.  In
particular, we must avoid exponential increases in the amount of
resource usage (either time or space) --- the implementation must be
\emph{scalable}\footnote{Blume-Kohout, Caves, and
Deutsch~\cite{Blume-Kohout02} give a general characterization of the
requirements for scalability.}.

%
Therefore, the importance of an experimental achievement of an early
milestone (such as the four-element Grover's algorithm) should be
measured primarily on these criteria.  A demonstration that Grover's
algorithm finds the marked item in fewer steps than is possible with a
classical computer is an important goal, but it is less important than
the fine level of quantum control that it implies.  At this early
stage of development of quantum computers, any such demonstration is a
significant achievement, while a demonstration of such control in a
scalable manner is likely to be significantly more difficult and
consequently more impressive.

This is illustrated by the experiment of Kwiat \etal~\cite{Kwiat00},
which demonstrated the ability to implement the search algorithm in a
quantum optical system, but using an encoding that is not scalable ---
as they point out, the number of optical elements that they require
grows exponentially in the number of qubits in their system.  Thus,
although their techniques might be successfully extended to a few
qubits, they are not practical as the basis for an approach to
building a quantum computer.

%
In contrast, we are explicitly concerned with developing experimental
milestones on the path toward full-scale quantum computation in
optical systems.  We show that Grover's algorithm on four elements
provides several experiments that gradually increase in complexity.
The simplest version requires little more than a single,
coincidence-basis $\cnot$ gate together with a source of entangled
photon pairs, while the most complex version requires two scalable
$\cnot$ gates and six photons.

%
Before describing these experiments and their requirements, we give a
brief description of Grover's algorithm
(Section~\ref{se:grover's_algorithm}) and LOQC
(Section~\ref{se:LOQC_with_polarization_encoding}).  Since the
original proposal of LOQC, there have been many simplifications and
improvements to the scheme.  We give a concise description of the
basics of LOQC making full use of these simplifications, focusing on a
variant of the original scheme that uses polarization-encoded qubits.
In Sections~\ref{se:the_two_qubit_grover_in_LOQC}
and~\ref{se:simplifications}, we describe and compare several optical
circuits, all implementing Grover's algorithm on four elements.  In
Section~\ref{se:figures_of_merit} we briefly discuss appropriate
figures of merit for Grover's algorithm, and we conclude in
Section~\ref{se:a_hierarchy_of_experiments}.

%
\section{Grover's algorithm on four elements}
\label{se:grover's_algorithm}
%
Grover's algorithm~\cite{Grover96,Grover97} (see also Nielsen and
Chuang~\cite{Nielsen00} for an elementary treatment on which much of
this section is based) is a quantum algorithm that can speed up the
solution to certain types of oracle-based computations.  We will say
more about oracles and their implementation after describing Grover's
algorithm.

\subsection{Grover's algorithm}

%
Suppose our search space consists of $N\eq2^n$ elements, of which one
is a solution to a given problem.  Grover's algorithm identifies the
solution (with high probability) using $n+1$ qubits according to the
following algorithm:
\begin{enumerate}
\item Prepare the state $|0\ket^{\tp n}|1\ket$.
\item Apply $R^{\tp n+1}$, where
$R=\frac{1}{\sqrt2}\[\begin{smallmatrix}1&1\\1&-1\end{smallmatrix}\]$
is the one-qubit Hadamard gate.  (We use the symbol $R$ instead of the
usual $H$ to avoid confusion with the horizontal polarization state.)
\item Apply the oracle, which flips the ancilla qubit conditional on
the other qubits being in the state corresponding to the solution.
\item Apply $R^{\tp n+1}$.
\item Apply a phase shift to the data qubits conditional on not being
in the state $|0\ket^{\tp n}$, described by the unitary operator
$2|0\ket\bra0|^{\tp n}-I_n$ where $I_n$ is the identity operation on
the data qubits.
\item Apply $R^{\tp n}$.
\item Repeat steps (3) to (6) a specified number of times, then
measure the qubits in the computational basis.
\end{enumerate}
The number of repetitions (which is also the number of uses of the
oracle) that maximizes the probability of obtaining the correct answer
is the nearest integer to
\begin{equation} \label{eq:number_iterations}
\frac{\arccos\sqrt{1/N}}{2\arccos\sqrt{(N-1)/N}}
\end{equation}
(Boyer \etal~\cite{Boyer98}, see also~\cite{Nielsen00}).  This number
is bounded above by $\lceil\pi\sqrt{N}/4\rceil$, hence the claim that
Grover's algorithm uses $\mathcal O(\sqrt N)$ oracle calls, compared
to the $\mathcal O(N)$ oracle calls required in the classical case.
%
%
\begin{figure}[t]
\scalebox{0.9}{\includegraphics{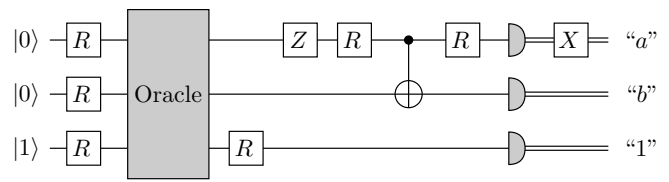}}
\caption{\label{fi:four_element_grover} A circuit diagram for the
four-element Grover algorithm, based on the figure in Box 6.1
of~\cite{Nielsen00}.  The top two qubits are the data qubits,
initialized in state $|0\ket|0\ket$, while the bottom qubit is the
ancilla qubit, initialized in state $|1\ket$.  The boxes labeled $R$
and $Z$ represent the one-qubit Hadamard and Pauli $\sigma_Z$ gates,
respectively.  The $\cnot$ is denoted by the usual symbol, while the
gray half-circles represent one-qubit measurements in the
computational basis, whose output appears on the classical output
wires (double lines).  The final $X$ gate represents the classical
{\sc not} required to put the output into the correct form.  The
measurement always gives ``1'' on the ancilla qubit, while the data
qubits give ``$a$'' and ``$b$''.  It is straightforward to show that,
in principle, $ab$ is the state marked by the oracle.}
\end{figure}

%
%
For the remainder of this paper, we restrict our attention to the case
where the number of elements in the search space is $N=4$.  In that
case, the number of repetitions specified by
Eq.\,(\ref{eq:number_iterations}) is exactly one.  A simplified
circuit based on the algorithm described above is shown in
Fig.\,\ref{fi:four_element_grover}.  It can be verified directly that
this circuit, using only one oracle call, gives the correct answer
with probability 1, compared to the average of 2.25 oracle calls that
must be made with a classical circuit.  For example, if the solution
is 10, then the output of the circuit is $a=1$ and $b=0$.

\subsection{Implementing the oracle}

An oracle is a quantum circuit that \emph{recognizes} solutions to a
given problem.  For example, suppose we wish to solve a version of the
traveling salesman problem, where the goal is to find a route visiting
a given collection of cities that is shorter than some specified
length $L$.  Although it is in general hard to find such a route, it
is easy to recognize whether a proposed route solves the problem:
simply add up the total distance the salesman would travel on the
proposed route, and compare it to $L$.

%
Specifically, an oracle is a circuit that, given an input consisting
of a potential solution to a problem, flips the sign of an ancilla
qubit if and only if the input is a solution to the problem.  Since
the only action of the oracle is to recognize solutions, its internal
structure is unimportant in a test of the algorithm itself.  Thus, for
our purposes, the choice of oracle is arbitrary, and may be chosen to
be as simple as possible.

\begin{figure}[t]
\scalebox{0.9}{\includegraphics{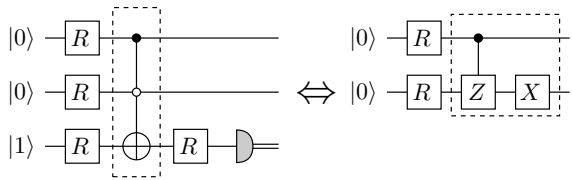}}
\caption{\label{fi:oracle} The circuit on the left shows the beginning
of the Grover circuit with an example oracle (inside the dashed box)
marking the item 10.  We have implemented the oracle using a variant
of the Toffoli gate, where the state of the third qubit is flipped
when the first two qubits are in the state $|10\ket$, as indicated by
the closed and open circles on the control qubits.  We have moved the
measurement on the third qubit forward since it plays no further role
in the algorithm.  In the text, we show that this circuit is in fact
equivalent to the simplification on the right, where the Toffoli has
been replaced by a controlled-$Z$ ($\csign$) operation followed by an
$X$ on the appropriate qubit.}
\end{figure}

%
Although the internal workings of the oracle are unimportant for the
purposes of testing the algorithm, the complexity of implementing
\emph{some} oracle must be included to characterize the difficulty of
performing the experiment.  A simple implementation of an oracle
marking one of four states is a Toffoli gate, with the control qubits
negated where necessary to specify any of the states 00, 01, 10, or 11
(see the left-hand side of Fig.\,\ref{fi:oracle} for the example where
the marked state is 10).
%
%

If the marked state is 10, the action of the oracle on the three
qubits is to take the state
$(|00\ket+|01\ket+|10\ket+|11\ket)(|0\ket-|1\ket)$ to
\begin{equation}\begin{split}
(|00\ket+|01\ket+|11\ket)(|0\ket-|1\ket)+|10\ket(|1\ket-|0\ket)\\
=(|00\ket+|01\ket-|10\ket+|11\ket)(|0\ket-|1\ket)
\end{split}\end{equation}
(omitting the normalization).  Thus the oracle simply has the effect
of flipping the sign of the marked state.  The ancilla is not used
again, so it can be discarded at this point.

Toffoli gates are difficult to implement in LOQC because there is no
known way to implement one without using several $\cnot$s.  However,
for our purposes, a full Toffoli gate is not required because the
ancilla qubit plays such a limited role.  The two-qubit circuit on the
right-hand side of Fig.~\ref{fi:oracle} illustrates this for the case
where the marked state is 10.  A single controlled-$Z$ ($\csign$) gate
that flips the sign of the $|11\ket$ state, followed by $X$ gates to
move the minus sign to the appropriate state, has the same action as
the original oracle.

\begin{figure}[t]
\scalebox{0.9}{\includegraphics{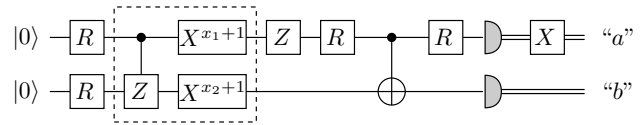}}
\caption{\label{fi:four_element_grover_with_oracle} Inserting the
simplified oracle of Fig.~\ref{fi:oracle} into the circuit of
Fig.~\ref{fi:four_element_grover} gives this circuit.  Note that the
marked state is specified inside the oracle (the dashed box) by the
values of $x_1$ and $x_2$ used to determine whether or not the $X$
gates are applied.  (Note that addition in the exponent of the $X$
gates is modulo 2.)  Under ideal circumstances, the output of the
circuit is $a=x_1$ and $b=x_2.$}
\end{figure}

A simplified circuit to implement the four-element Grover algorithm is
given in Fig.\,\ref{fi:four_element_grover_with_oracle}.  This is the
circuit that we will work with for the remainder of this paper.
%
%

%
\section{LOQC with polarization encoding}
\label{se:LOQC_with_polarization_encoding}

%
In LOQC, qubits are encoded in \emph{dual rail logic}~\cite{Knill01b}:
Two modes $A$ and $B$ are used, and logical $|0\ket$ and $|1\ket$ are
encoded as $|1\ket_A|0\ket_B$ and $|0\ket_A|1\ket_B$, respectively.
The modes may represent two different \emph{spatial} modes, or two
different \emph{polarization} modes of a single spatial
mode.\footnote{A good introduction to LOQC in spatial encoding is
provided by a set of lectures of Knill, available online at {\tt
http://online.itp.ucsb.edu/online/qinfo01/}.}

In practice, it is likely that polarization-encoded qubits will be
used, so that logical $|0\ket$ and $|1\ket$ are encoded as $|H\ket$
and $|V\ket$, respectively, where $H$ and $V$ refer to horizontal and
vertical polarization one-photon states of the same spatial mode.  The
main reasons for this are (1) it significantly simplifies the
implementation of the $\cnot$ gate (see below), (2) it allows
one-qubit gates to be implemented using only waveplates and phase
delays rather than beamsplitters and interferometers, and (3) it
reduces the effects of noise by ensuring that, unlike with spatial
encoding, both states follow the same path on the quantum wires
between gates.  In this section we describe in some detail the
construction of one-qubit gates and $\cnot$s in polarization-encoded
LOQC.

\subsection{One-qubit gates}

%
%
\begin{table}[b]
\begin{tabular}{|l|l|}
\hline
Gate        & Optical element                                       \\
\hline
$e^{i\theta}I=e^{i\theta}\[\begin{smallmatrix}1&0\\0&1\end{smallmatrix}\]$
            & phase delay of $-\theta$                              \\
$R=\frac{1}{\sqrt2}\[\begin{smallmatrix}1&1\\1&-1\end{smallmatrix}\]$
            & waveplate with $\phi=180^\circ$, $\alpha=-67.5^\circ$ \\
$T=\[\begin{smallmatrix}1&0\\0&e^{i\pi/4}\end{smallmatrix}\]$
            & waveplate with $\phi=45^\circ$, $\alpha=90^\circ$     \\
$X=\[\begin{smallmatrix}0&1\\1&0\end{smallmatrix}\]$
            & waveplate with $\phi=180^\circ$, $\alpha=-45^\circ$   \\
$Z=\[\begin{smallmatrix}1&0\\0&-1\end{smallmatrix}\]$
            & waveplate with $\phi=180^\circ$, $\alpha=90^\circ$    \\
$Y=\[\begin{smallmatrix}0&-i\\i&0\end{smallmatrix}\]$
            & 2  waveplates and a phase delay: $(e^{i\pi/4}I)XZ$    \\
\hline
\end{tabular}.
\caption{\label{ta:one-qubit_gates} Various one-qubit gates and their
implementation in polarization-encoded LOQC.  $\alpha$ and $\phi$
refer to the angle of the slow axis to the horizontal and the relative
phase added to light parallel to the slow axis, respectively.  Note
that $T$ requires a waveplate with a relative delay of one eighth of a
wavelength.}
\end{table}
To our knowledge, no complete description of how to implement basic
quantum gates with polarization encoding has been given in the
literature, so we provide one here.  For one-qubit gates, waveplates
and phase delays are sufficient.  A waveplate with slow axis $|H'\ket$
and fast axis $|V'\ket$ has action
\begin{equation}
\begin{split}
|H'\ket&\to e^{i\phi}|H'\ket\\
|V'\ket&\to|V'\ket,
\end{split}
\end{equation}
where $\phi$ is the resulting relative phase difference.  Special
cases in common use are the half and quarter waveplates, with $\phi$
equal to half and a quarter of a wavelength, respectively.  Now
suppose $|H'\ket$ is rotated counter-clockwise (with respect to the
direction of travel of the light) by an angle $\alpha$ from $|H\ket$.
If the input state is $\[\begin{smallmatrix}h\\v\end{smallmatrix}\]\eq
h|H\ket+v|V\ket$, then the output is given by
\begin{equation} \label{eq:waveplate}
\[\begin{matrix}e^{i\phi}\cos^2\alpha+\sin^2\alpha&(e^{i\phi}-1)\cos\alpha\sin\alpha\\(e^{i\phi}-1)\cos\alpha\sin\alpha&e^{i\phi}\sin^2\alpha+\cos^2\alpha
\end{matrix}\]\[\begin{matrix}h\\v\end{matrix}\].
\end{equation}

Special cases of this transformation for common one-qubit gates are
set out in Table~\ref{ta:one-qubit_gates}.  The Hadamard and $\pi/8$
gates, labeled $R$ and $T$ in the table, are a universal set for
one-qubit quantum computation (Boykin \etal~\cite{Boykin00}), and so
any one-qubit gate can be obtained by a sequence of waveplates,
although it is convenient to allow phase delays as well.  In the
Grover circuit, the only one-qubit gates used are the $R$, $X$, and
$Z$ gates, and thus we only require half waveplates.

%
%
\begin{figure*}[t]
\scalebox{1.0}{\includegraphics{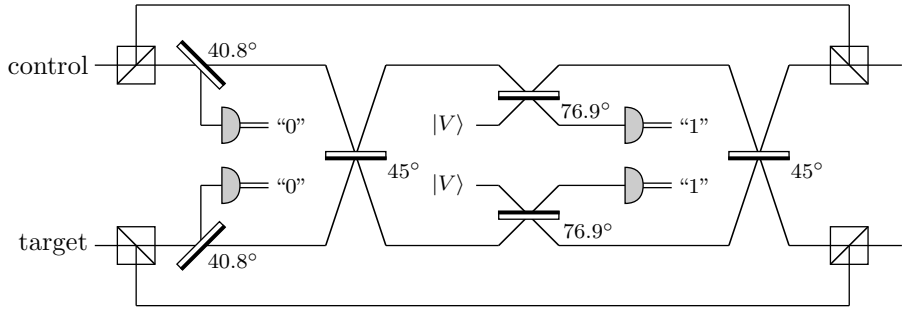}}
\caption{\label{fi:knill_csign} The simplified KLM $\csign$
of~\cite{Ralph01}.  The top rail contains the control qubit and the
bottom rail contains the target qubit, both encoded in the
polarization of a single photon.  A square with a diagonal line across
it represents a polarizing beamsplitter.  By convention, we always
assume that the horizontal polarization is 100\% reflected while the
vertical polarization is 100\% transmitted.  So, for example, after
the first polarizing beamsplitters, the topmost rail contains the
horizontally polarized component of the control qubit.  A thin
rectangle represents an ordinary beamsplitter, with a sign change for
the mode reflected from the thick black side and reflectivity given by
the cosine of the angle written next to it.  (If the input modes to a
beamsplitter are $|a\ket_\text{in}$ and $|b\ket_\text{in}$, with the
$b$ mode receiving the sign change and with reflectivity given by
$\cos x$, then the outputs are $\cos x|a\ket_\text{out}+\sin
x|b\ket_\text{out}$ and $\sin x|a\ket_\text{out}-\cos
x|b\ket_\text{out}$.)  The circuit uses two vertically polarized
ancilla photons.  It succeeds if the first two measurements both count
0 photons and the second two measurements both count 1 photon.}
\end{figure*}
\subsection{Two-qubit gates}

Since the publication of the original LOQC scheme of
KLM~\cite{Knill01b}, many simplifications of their $\csign$ gate have
been developed, with varying trade-offs between simplicity and
functionality.  The different types may be divided into two classes,
those that are scalable and those that are not.  In this section, we
describe both types.  (Note that the $\cnot$ and $\csign$ gates are
related by conjugation by Hadamards on the target bit, \ie
$\cnot=(I\tp R)\csign(I\tp R)$.  Therefore, in the context of LOQC
where one-qubit gates are relatively straightforward, these two gates
are practically equivalent, and we will use the two almost
interchangeably.)

\subsubsection{Scalable two-qubit gates}

%
%
\begin{figure*}[t]
\scalebox{1.0}{\includegraphics{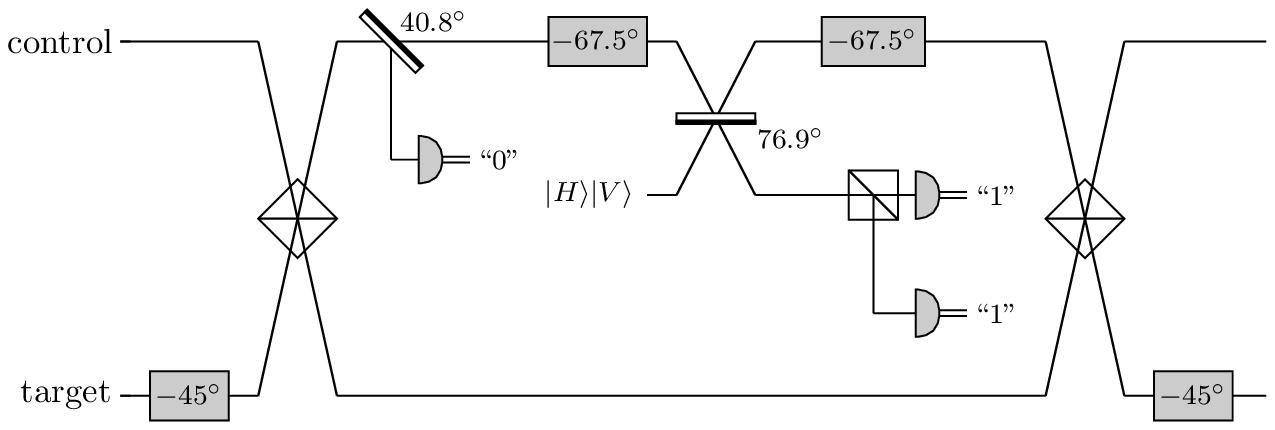}}
\caption{\label{fi:pol_csign} A further simplified (but still
scalable) polarization-encoded KLM $\csign$.  A gray rectangle
containing ``$x^\circ$'' represents a half waveplate with slow axis at
an angle of $x^\circ$ to the horizontal polarization.  See
Table~\ref{ta:one-qubit_gates} for the corresponding one-qubit gates.
This circuit works similarly to the previous one
(Fig.~\ref{fi:knill_csign}), but it takes fuller advantage of the
orthogonality of the polarization states.  For a full description, see
the text.}
\end{figure*}
The KLM scheme~\cite{Knill01b} has two properties that at first appear
contradictory: the LOQC $\csign$ is non-deterministic, but it is used
to do computations in a scalable manner.  The non-deterministic nature
of the KLM $\csign$ is essential to engineer a two-photon interaction
without using highly nonlinear materials, but it poses a problem: if
its success probability is $\epsilon<1$, then the success probability
of a circuit with $n$ $\csign$s is $\epsilon^n$ --- it decreases
exponentially with $n$.  A solution to this problem is the technique
of \emph{gate teleportation} described by Nielsen and
Chuang~\cite{Nielsen97c} and Gottesman and Chuang~\cite{Gottesman99}.
This technique allows the gates to be prepared as an offline resource,
and then ``teleported in'' whenever required for a computation.  KLM
showed that the teleportation step can be made near-deterministic
using a sufficiently large number of repetitions.  This technique is
unlikely to be used in early experiments, however, because the extra
difficulty involved in teleporting gates will more than cancel out the
advantages of increasing the success probability when the number of
$\csign$s is small.

An essential feature required to make this work is that it must be
possible to determine when the gate has succeeded.  The KLM $\cnot$
has this property --- although it only succeeds 1 time in 16, whether
or not it has succeeded is determined by the outcomes of measurements
on ancilla photons.  We use the term \emph{scalable} to describe a
$\csign$ (or $\cnot$) that has the property that it is known when it
succeeds.

In this paper, we will not work directly with the KLM $\csign$ since
there are simpler alternatives, such as the closely related
simplification proposed by Ralph \etal~\cite{Ralph01} and the
substantial modification proposed by Knill~\cite{Knill02}.  There is
also a promising alternative approach using entangled ancillas
discovered by Pittman, Jacobs and Franson~\cite{Pittman01} that we
will not consider further here.  We focus on the $\csign$ of Ralph
\etal, shown in Fig.~\ref{fi:knill_csign}.\footnote{Recent numerical
work by Lund, Bell, and Ralph~\cite{Lund03}, shows that the simplified
KLM $\csign$ of~\cite{Ralph01} is more resilient to detector and
ancilla inefficiencies than the other two, perhaps because it acts
symmetrically on the two qubits.  For example, the fidelity of this
gate (calculated as the fidelity of the actual output with the ideal
output, minimized over input states) is larger than the fidelities of
the other two gates for detector efficiencies up to approximately
95\%.  However, it remains to be seen what effects other sources of
error, such as mode-matching errors, and imperfect beamsplitter
reflectivities, will have on the relative merits of each gate.}

In fact, there is a further, substantial simplification to this
circuit that is achieved by making fuller use of the polarization
encoding, resulting in the circuit in Fig.~\ref{fi:pol_csign}.  This
gate still requires two ancilla photons.  However, it uses fewer
detectors, beamsplitters, and polarizing beamsplitters, and eliminates
two interferometers.  Its effect on qubit states is unchanged, up to
an unimportant overall phase of $-1$.  If we denote the beamsplitter
reflectivities as $\eta_1\eq5-3\sqrt2$ and $\eta_2\eq(3-\sqrt2)/7$
(which are approximated as $\cos40.8^\circ$ and $\cos76.9^\circ$ in
the diagram), then the action of the gate is the following:
\begin{equation}
\begin{split}
&|00\ket\to\sqrt{\eta_1\eta_2}(2\eta_2-1)|00\ket=-\sqrt p|00\ket \\
&|01\ket\to\eta_1(3\eta_2^2-2\eta_2)|01\ket=-\sqrt p|01\ket \\
&|10\ket\to\sqrt{\eta_1\eta_2}(2\eta_2-1)|00\ket=-\sqrt p|10\ket \\
&|11\ket\to\eta_2|11\ket=\sqrt p|11\ket \\
\end{split}
\end{equation}
where the success probability $p$ is given by
$p\eq\eta_2^2=(11-6\sqrt2)/49\approx0.05$.  Thus the gate works
approximately 1 out of every 20 attempts.  For the remainder of this
paper, we will refer to this gate simply as a ``scalable $\csign$''.

\subsubsection{Coincidence-basis two-qubit gates}

%
%
\begin{figure}[t]
\scalebox{1.0}{\includegraphics{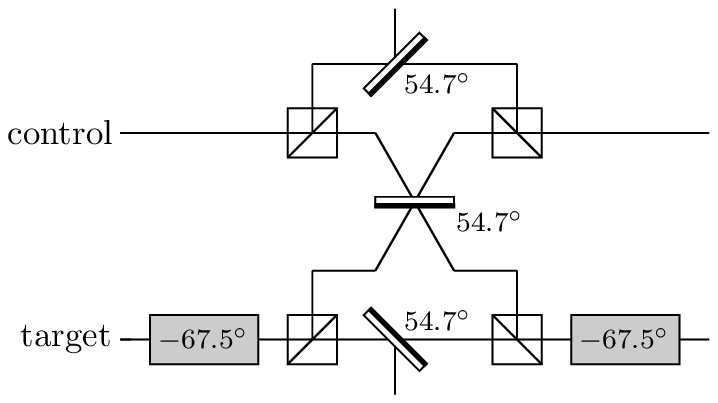}}
\caption{\label{fi:ralph_cnot} The coincidence-basis $\cnot$
of~\cite{Hofmann02,Ralph02}.  All three beamsplitters have the same
reflectivity $1/3\approx\cos54.7^\circ$.  It can be turned into a
$\csign$ by removing the two halfwave plates.  Note that it is not
necessary to have detectors on the reflected modes of the topmost and
bottommost beamsplitters (even though measuring a photon in either of
these modes would signal a failure), since other failures of this gate
are undetectable until the end of the computation.  The gate has
worked if exactly one photon is found in each rail.}
\end{figure}
An even simpler, but non-scalable $\cnot$ was discovered by Hofmann
and Takeuchi~\cite{Hofmann02} and Ralph \etal~\cite{Ralph02}.  It
succeeds 1 time in 9, but it only works in the \emph{coincidence
basis}, \ie when the results of the whole computation are selected to
contain an allowed distribution of photons among detectors.  We call
this a \emph{coincidence-basis $\cnot$}.  See
Fig.~\ref{fi:ralph_cnot}.  This circuit has been designed so that if
exactly one photon is measured in the top rail (in either
polarization), and one in the bottom rail, it has worked with
certainty.  Otherwise, the result is discarded and the experiment is
repeated.  It cannot in general be followed by further two-qubit
gates, as it is possible for a later gate to mask a failure.  Thus it
cannot be used to do scalable quantum computation.

%
The useful purpose served by this gate (as well as the
coincidence-basis gate of~\cite{Pittman03}) is as a simpler
intermediate step before the full complexity of a scalable $\cnot$.
In a general circuit, it may be possible to replace one or more
scalable $\cnot$s with a coincidence-basis $\cnot$, thereby
significantly reducing the complexity of circuits containing a few
$\cnot$s.  In the following sections on constructing optical circuits
to perform the four-element Grover algorithm, we will see some of
these ideas in action.

%
\section{The two-qubit Grover in LOQC}
\label{se:the_two_qubit_grover_in_LOQC}
%
%
\begin{figure*}[t]
\scalebox{0.78}{\includegraphics{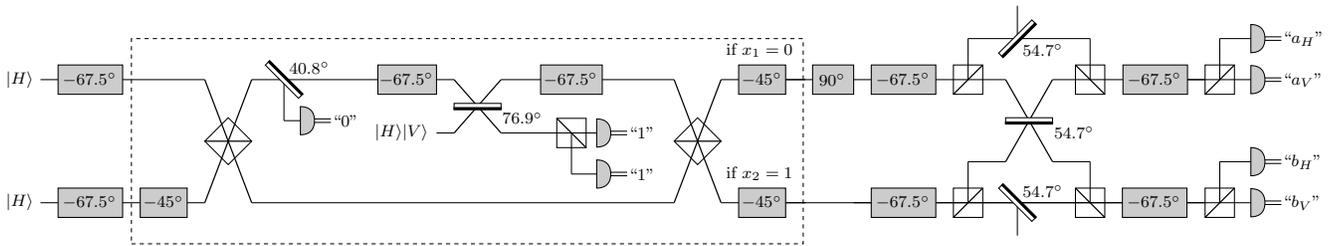}}
\caption{\label{fi:optical_grover} An optical implementation of
Grover's algorithm on four elements based on the circuit in
Fig.~\ref{fi:four_element_grover_with_oracle}.  The oracle part is
contained in the dashed box.  This circuit is essentially the
concatenation of the circuits for the scalable $\csign$
(Fig.~\ref{fi:pol_csign}) and the coincidence-basis $\cnot$
(Fig.~\ref{fi:ralph_cnot}), together with a few extra waveplates.  The
output of the circuit is discarded unless the first measurement counts
0 photons, the second two measurements both count 1 photon, and one
photon is found in each pair of detectors at the end, \ie
$a_H+a_V=b_H+b_V=1$.  Note that we have omitted the final correcting
{\sc not} gate on the classical output in this diagram, but it should
still be done.  For example, if the oracle marks state 10, then the
algorithm has successfully identified the marked state if measurements
return $a_H=1$, $a_V=0$, $b_H=1$, $b_V=0$.}
\end{figure*}
%
%
A simplified circuit for the four-element Grover algorithm was given
in Fig.\,\ref{fi:four_element_grover_with_oracle}.  In
Fig.\,\ref{fi:optical_grover}, this circuit is translated directly
into an optical circuit, using the prescriptions and circuits of the
previous section.

%
The circuit, which succeeds one time in approximately $180=20\times9$
(the product of the number of attempts per success for each $\cnot$),
uses 10--12 half waveplates,\footnote{Note that the $90^\circ$ and
$67.5^\circ$ halfwave plates cannot be combined into a single
waveplate: their product
$\frac{1}{\sqrt2}\[\begin{smallmatrix}1&-1\\1&1\end{smallmatrix}\]$
has terms of opposite sign in the off-diagonal terms, while the
waveplate equation (Eq.~(\ref{eq:waveplate})) has these entries
equal.} 5 beamsplitters (2 of which must be modematched), 9 polarizing
beamsplitters (4 of which must be modematched), 4 photons that must be
simultaneously produced in desired polarization states, and 7
single-photon detectors.  The second $\cnot$ can be done in the
coincidence basis since there are no interactions between the two
qubits following it.  Therefore, if the final measurement contains an
allowed distribution of photons (exactly one in the top two detectors
and one in the bottom two detectors), we know that the second $\cnot$
worked, which is sufficient for our purposes here.\footnote{A small
but potentially useful simplification is to remove the $40.8^\circ$
beamsplitter, as described in~\cite{Lund03}.  They show that, until
detector and source efficiencies of up to approximately 99.5\% are
reached, the fidelity of the gate can be substantially increased by
removing this beamsplitter and adjusting the reflectivity of the
$76.9^\circ$ beamsplitter.  Given that beamsplitter reflectivities are
imperfect, removing this beamsplitter is likely to decrease that
source of error, while also decreasing the complexity of the circuit
by removing a detector.  There is a catch, however: the probability of
success decreases by a factor of 4--5 for efficiencies of 80\%--95\%.}

However, it is important to note that the output of this circuit
(before the measurement) could not be used to do further calculations
because of the uncertainty in the outcome of the second $\cnot$.  If,
for example, there were two photons in the top rail after the second
$\cnot$, the system's state would no longer be in the ``qubit space''.
A third $\cnot$ might bring the system back into the qubit space, but
it is unlikely to have performed the transformation we expected.  In
this case, the overall circuit fails, but we have no way of detecting
the failure (except to compare with the answer that we can calculate
by hand for this simple case).

To ensure reliability for further calculations, the second $\cnot$
should be replaced by a scalable $\cnot$.  The optical circuit for
this case would work one time in 400, and would contain on the order
of 14--16 waveplates, 8 polarizing beamsplitters (6 of which would be
modematched), 4 ordinary beamsplitters (2 of which would be
modematched), 6 photons produced in desired polarization states
simultaneously, and 10 single-photon detectors.  This would be
considerably more difficult to achieve experimentally.  Since we are
(in principle) guaranteed to be in the qubit space at the end of this
circuit, the output of each pair of detectors should contain exactly
one photon.  Therefore, it is possible to simplify the final detection
process by simply blocking out one of the polarizations (horizontal,
say), and then looking to see if a photon is detected.  This would
reduce the number of polarizing beamsplitters to 6 and the number of
detectors to 8, at the cost of introducing two polarization filters.
However, in practice the number of photons at the output will
sometimes be incorrect.  Thus, the increase in simplicity would have
to be weighed against the failures that would go undetected.

%
\section{Simplifications}
\label{se:simplifications}
\begin{figure*}
\scalebox{1.0}{\includegraphics{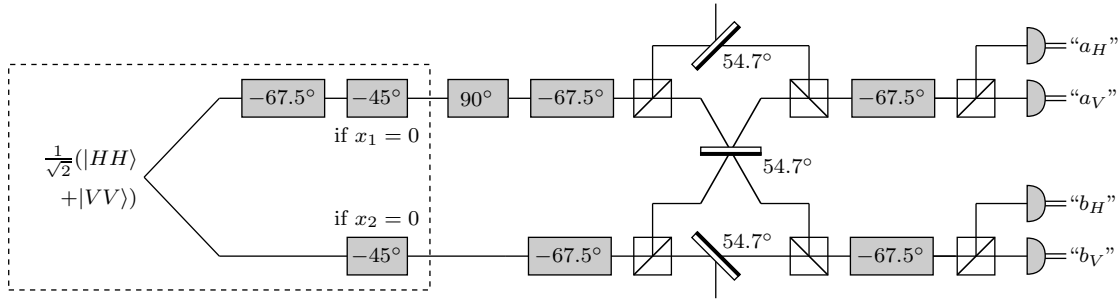}}
\caption{\label{fi:bell_state_grover} Grover's algorithm using a
parametric down conversion input.  This circuit works similarly to the
previous one, but the oracle is no longer demarcated from the initial
part of the circuit.  The dashed box in this figure now contains both
the oracle and the initialization to the state $|HH\ket+|VV\ket$.  The
advantage of this circuit is that it makes use of a natural source of
optical entanglement (parametric down conversion) to replace the very
difficult scalable $\csign$.  The outputs from this circuit are
accepted under the same conditions as the previous circuit
($a_H+a_V=b_H+b_V=1$), and the final classical {\sc not} has again
been omitted.}
\end{figure*}

%
By far the most difficult aspect of the experiments just described is
implementing the scalable $\csign$s.  However, the $\csign$ in the
oracle is only used in a very restricted way, and it turns out that we
can replace it with a much simpler circuit.  Since only one input
state is ever used, namely $(|H\ket+|V\ket)(|H\ket+|V\ket)$, only one
state is ever output from the $\csign$, namely
$|HH\ket+|VH\ket+|HV\ket-|VV\ket$.  (We will continue to neglect
normalization constants.)  If a source of entangled input states were
available, then the $\csign$ could be replaced.  In optics, such a
source is in fact readily available: a parametric down conversion
source can be used to produce the state $|HH\ket+|VV\ket$, which can
be converted into our desired state by a Hadamard gate on the first
qubit, $|H\ket\to|H\ket+|V\ket$, $|V\ket\to|H\ket-|V\ket$.  Using this
fact, a much simplified version of Grover's algorithm is presented in
Fig.\,\ref{fi:bell_state_grover}.

%
%

%
The simplicity of this circuit compared with the previous one is
emphasized by comparing the number of components.  This circuit works
one in every 9 attempts, and requires 6--8 waveplates, 6 polarizing
beamsplitters (of which 2 must be modematched), 3 ordinary
beamsplitters (1 of which must be modematched), 2 photons which are
produced as the output of a parametric down-conversion source, and 4
single-photon detectors.

%
What have we traded for this enormous gain in simplicity?  It turns
out that we have compromised the versatility of the algorithm.  Most
significantly, the oracle is no longer easily replaceable.  In
principle, the oracle should be a `''plug-in'' component able to have
many different forms corresponding to different potential problems.
In this simplified scheme, however, we have obscured the line between
the oracle and non-oracle parts of the circuit, making it difficult to
see how to make the circuit solve a problem using a different oracle.
In Fig.\,\ref{fi:bell_state_grover}, a dashed box outlines the
``oracle'' part of the circuit for comparison with the previous
diagrams, but there is in fact no clear line dividing the oracle from
the earlier part of the circuit.

This change affects how the circuit could be used.  One example is
demonstrating the variation in the success probability of Grover's
algorithm as a function of the number repetitions of steps (3)--(6)
described in Sec.\,\ref{se:grover's_algorithm}.  In the circuit in
Fig.\,\ref{fi:optical_grover}, the oracle can be reused with some
small changes.\footnote{The oracle on the right-hand side of
Fig.~\ref{fi:oracle} is designed to work with inputs that are equal
superpositions of computational basis states.  If the oracle is used
twice in the same circuit, then it is unlikely that the input state
will always be the same.  In order to make the oracle work for an
arbitrary input state, it is necessary to simply duplicate the $X$
gates following the $\csign$, before the $\csign$.  For the example in
Fig.~\ref{fi:oracle}, where the oracle marks the state $|01\ket$, the
oracle should consist of the following: an $X$ gate acting on the
bottom qubit, followed by the $\csign$, followed by the $X$ acting on
the bottom qubit.}  On the other hand, in
Fig.\,\ref{fi:bell_state_grover}, this is not possible --- the
``oracle'' can only be used once\footnote{For a more speculative
example, Grover's algorithm can be used to obtain upper bounds on an
entanglement monotone called the \emph{Groverian entanglement}, as
described by Biham, Osborne and Nielsen~\cite{Biham02}.  The basic
idea is that if an $n$-qubit state $\rho$ (possibly mixed) is used as
input rather than $|0\ket^{\otimes n}$, the square root of 1 minus the
success probability gives a good measure of the entanglement of
$\rho$.  This application requires input states with varying degrees
of entanglement, and thus is not possible in the simplified circuit.}.

%
\section{Figures of merit}
\label{se:figures_of_merit}
An important question that has so far not been addressed is what the
appropriate figures of merit are for this experiment.  There are two
related but distinct notions of success here.  The first is to what
extent the actual goal of Grover's algorithm has been achieved, \ie
how successfully the experiment distinguishes between the four
different oracles.  The second is how similar the actual operation of
circuit is to the ideal operation.  This second notion is important
for using these experiments as tests of the ability to combine the basic
elements of quantum computation.  It is clearly related to the first
--- if the experiment cannot reliably distinguish between the oracles,
then the actual behavior of the circuit must be very far from the
ideal operation.

In order to be able to compare experiments (and also to optimize the
performance of a particular experimental setup), we need to be more
precise about how to measure the success of these experiments.  We
suggest calculating figures of merit reflecting each of the two
notions of success described above.  The first is simply to measure
the distinguishability of the distribution of measurement results
output by the circuit for different oracles.  For example, suppose
that for the oracle marking the state $00$, the results $00$, $01$,
$10$, and $11$ occur with probabilities
$p_{00}\eq\{0.9,0.04,0.02,0.04\}$, while the corresponding results
when the oracle marks state $10$ are
$p_{10}\eq\{0.01,0.08,0.8,0.11\}$.  A simple indicator of the
distinguishability of these two distributions is their fidelity
\begin{equation}
F(p_{00},p_{10})\eq\sum_x\sqrt{p_{00}(x)p_{10}(x)}
\end{equation}
where $x$ ranges over the measurement outcomes $00,\dots,11$ and
$p_{ab}(x)$ is the probabilities of obtaining result $x$ given that
the oracle marked state $ab$.  This quantity has the property that it
is 1 precisely when the two distributions are identical and 0
precisely when the two distributions are non-overlapping, that is,
when the set of results for which the first distribution is non-zero
has no elements in common with the set of results for which the second
distribution is non-zero.

In the context of Grover's algorithm, it is desirable to make the
fidelity between the distributions arising from each pair of oracles
as large as possible.  (For an introduction to the fidelity, see, for
example,~\cite{Fuchs96, Nielsen00}.  The relationship of the fidelity
to distinguishability is explored in Wootters~\cite{Wootters81}
and~\cite{Fuchs96}.)

The second figure of merit is related to the similarity of the actual
operation implemented $\E$ to the desired unitary $U$.  $U$ is
obtained by simply multiplying together the circuit elements in
Fig.~\ref{fi:four_element_grover_with_oracle}.  $\E$, on the other
hand, must be determined experimentally.  Ideally, $\E$ should be
determined precisely using a method such as quantum process tomography
(Chuang and Nielsen~\cite{Chuang97} and Poyatos, Cirac, and
Zoller~\cite{Poyatos97}).  Although process tomography can be done
using only product-state inputs and one-qubit measurements, it
requires an enormous number of runs of the experiment since the output
states resulting from 16 different input density matrices must be
determined via quantum state tomography.

A less stringent, but much more easily calculated, criterion is that
the probability distributions for each oracle should be close to the
ideal distributions.  Again, it is desirable to have the fidelity of
the actual distribution to the ideal distribution for each oracle as
close to 1 as possible.\footnote{Knill \etal~\cite{Knill01c} have a
useful discussion of these issues where they advocate the
\emph{entanglement fidlelity} to measure the quality of an
experimental implementation of the five-qubit code.  They describe a
simple way of measuring the entanglement fidelity that could be easily
generalized to the setting of Grover's algorithm.}

%
\section{A hierarchy of experiments}
\label{se:a_hierarchy_of_experiments}
%
This collection of different implementations of the same algorithm
could be used as the basis for a series of experiments, each building
on the last, each more complicated than the last, each demonstrating
improved quantum control.  For example, once a basic coincidence-basis
$\cnot$ is working, it would be relatively simple to add a small
number of waveplates and a source of entangled photons to do the
circuit in Fig.\,\ref{fi:bell_state_grover}.  Once a scalable $\cnot$
is achieved, these two different $\cnot$ circuits could be combined to
do the more complicated implementation of Grover's algorithm in
Fig.\,\ref{fi:optical_grover}, demonstrating the ability to combine a
scalable $\cnot$ with further non-trivial quantum computations.
Finally, in the more distant future, the implementation using two
scalable $\cnot$s would make a good testing ground for techniques for
combining LOQC components.

%
\acknowledgments{JLD thanks Jeremy O'Brien, Geoff Pryde, and Andrew
White for providing insight into the world of experiments, Alexei
Gilchrist and Geoff Pryde for a careful reading of the manuscript,
Andrew Doherty for help in understanding waveplates, Michael Nielsen
for helpful discussions on Grover's algorithm, and Paul Cochrane and
Alexei Gilchrist for help using their program for drawing quantum
circuits ({\tt http://pyscript.sourceforge.net}), which produced all
of the diagrams in this paper.  JLD and GJM thank the Institute for
Quantum Information for their hospitality.  This work was supported in
part by the National Science Foundation under grant EIA--0086038, and
by the Australian Research Council and ARDA.}

\bibliographystyle{../../MODULES/REVTEX/apsrev}

\bibliography{../../MODULES/BIBLIOGRAPHY/mybib}

\begin{thebibliography}{28}
\expandafter\ifx\csname natexlab\endcsname\relax\def\natexlab#1{#1}\fi
\expandafter\ifx\csname bibnamefont\endcsname\relax
  \def\bibnamefont#1{#1}\fi
\expandafter\ifx\csname bibfnamefont\endcsname\relax
  \def\bibfnamefont#1{#1}\fi
\expandafter\ifx\csname citenamefont\endcsname\relax
  \def\citenamefont#1{#1}\fi
\expandafter\ifx\csname url\endcsname\relax
  \def\url#1{\texttt{#1}}\fi
\expandafter\ifx\csname urlprefix\endcsname\relax\def\urlprefix{URL }\fi
\providecommand{\bibinfo}[2]{#2}
\providecommand{\eprint}[2][]{\url{#2}}

\bibitem[{\citenamefont{Knill et~al.}(2001{\natexlab{a}})\citenamefont{Knill,
  Laflamme, and Milburn}}]{Knill01b}
\bibinfo{author}{\bibfnamefont{E.}~\bibnamefont{Knill}},
  \bibinfo{author}{\bibfnamefont{R.}~\bibnamefont{Laflamme}}, \bibnamefont{and}
  \bibinfo{author}{\bibfnamefont{G.~J.} \bibnamefont{Milburn}},
  \bibinfo{journal}{Nature (London)} \textbf{\bibinfo{volume}{409}},
  \bibinfo{pages}{46} (\bibinfo{year}{2001}{\natexlab{a}}),
  \bibinfo{note}{ar{X}iv:quant-ph/0006088}.

\bibitem[{QIS(2002)}]{QIST02}
 (\bibinfo{year}{2002}), \urlprefix\url{http://qist.lanl.gov}.

\bibitem[{\citenamefont{Grover}(1997)}]{Grover97}
\bibinfo{author}{\bibfnamefont{L.~K.} \bibnamefont{Grover}},
  \bibinfo{journal}{Phys.\ Rev.\ Lett.} \textbf{\bibinfo{volume}{79}},
  \bibinfo{pages}{325} (\bibinfo{year}{1997}),
  \bibinfo{note}{ar{X}iv:quant-ph/9706033}.

\bibitem[{\citenamefont{Grover}(1996)}]{Grover96}
\bibinfo{author}{\bibfnamefont{L.}~\bibnamefont{Grover}}, in
  \emph{\bibinfo{booktitle}{Proc.\ 28$^\text{th}$ Annual ACM Symposium on the
  Theory of Computation}} (\bibinfo{publisher}{ACM Press},
  \bibinfo{address}{New York}, \bibinfo{year}{1996}), pp.
  \bibinfo{pages}{212--219}, \bibinfo{note}{ar{X}iv:quant-ph/9605043}.

\bibitem[{\citenamefont{Chuang et~al.}(1998)\citenamefont{Chuang, Gershenfeld,
  and Kubinec}}]{Chuang98}
\bibinfo{author}{\bibfnamefont{I.~L.} \bibnamefont{Chuang}},
  \bibinfo{author}{\bibfnamefont{N.}~\bibnamefont{Gershenfeld}},
  \bibnamefont{and} \bibinfo{author}{\bibfnamefont{M.}~\bibnamefont{Kubinec}},
  \bibinfo{journal}{Phys.\ Rev.\ Lett.} \textbf{\bibinfo{volume}{18}},
  \bibinfo{pages}{3408} (\bibinfo{year}{1998}).

\bibitem[{\citenamefont{Jones et~al.}(1998)\citenamefont{Jones, Mosca, and
  Hansen}}]{Jones98b}
\bibinfo{author}{\bibfnamefont{J.~A.} \bibnamefont{Jones}},
  \bibinfo{author}{\bibfnamefont{M.}~\bibnamefont{Mosca}}, \bibnamefont{and}
  \bibinfo{author}{\bibfnamefont{R.~H.} \bibnamefont{Hansen}},
  \bibinfo{journal}{Nature (London)} \textbf{\bibinfo{volume}{393}},
  \bibinfo{pages}{344} (\bibinfo{year}{1998}),
  \bibinfo{note}{ar{X}iv:quant-ph/9805069}.

\bibitem[{\citenamefont{Gottesman et~al.}(2001)\citenamefont{Gottesman, Kitaev,
  and Preskill}}]{Gottesman00}
\bibinfo{author}{\bibfnamefont{D.}~\bibnamefont{Gottesman}},
  \bibinfo{author}{\bibfnamefont{A.}~\bibnamefont{Kitaev}}, \bibnamefont{and}
  \bibinfo{author}{\bibfnamefont{J.}~\bibnamefont{Preskill}},
  \bibinfo{journal}{Phys.\ Rev.\ A} \textbf{\bibinfo{volume}{64}},
  \bibinfo{pages}{012310} (\bibinfo{year}{2001}),
  \bibinfo{note}{ar{X}iv:quant-ph/0008040}.

\bibitem[{\citenamefont{Pittman et~al.}(2003)\citenamefont{Pittman, Fitch,
  Jacobs, and Franson}}]{Pittman03}
\bibinfo{author}{\bibfnamefont{T.~B.} \bibnamefont{Pittman}},
  \bibinfo{author}{\bibfnamefont{M.~J.} \bibnamefont{Fitch}},
  \bibinfo{author}{\bibfnamefont{B.~C.} \bibnamefont{Jacobs}},
  \bibnamefont{and} \bibinfo{author}{\bibfnamefont{J.~D.}
  \bibnamefont{Franson}}, \bibinfo{journal}{ar{X}iv:quant-ph/0303095}
  (\bibinfo{year}{2003}).

\bibitem[{\citenamefont{Kwiat et~al.}(2000)\citenamefont{Kwiat, Mitchell,
  Schwindt, and White}}]{Kwiat00}
\bibinfo{author}{\bibfnamefont{P.~G.} \bibnamefont{Kwiat}},
  \bibinfo{author}{\bibfnamefont{J.~R.} \bibnamefont{Mitchell}},
  \bibinfo{author}{\bibfnamefont{P.~D.~D.} \bibnamefont{Schwindt}},
  \bibnamefont{and} \bibinfo{author}{\bibfnamefont{A.~G.} \bibnamefont{White}},
  \bibinfo{journal}{Journal of Modern Optics} \textbf{\bibinfo{volume}{47}},
  \bibinfo{pages}{257} (\bibinfo{year}{2000}),
  \bibinfo{note}{ar{X}iv:quant-ph/9905086}.

\bibitem[{\citenamefont{Bhattacharya et~al.}(2002)\citenamefont{Bhattacharya,
  van~den Heuvell, and Spreeuw}}]{Bhattacharya02}
\bibinfo{author}{\bibfnamefont{N.}~\bibnamefont{Bhattacharya}},
  \bibinfo{author}{\bibfnamefont{H.~B.} \bibnamefont{van~den Heuvell}},
  \bibnamefont{and} \bibinfo{author}{\bibfnamefont{R.~J.~C.}
  \bibnamefont{Spreeuw}}, \bibinfo{journal}{Phys.\ Rev.\ Lett.}
  \textbf{\bibinfo{volume}{88}}, \bibinfo{pages}{137901}
  (\bibinfo{year}{2002}), \bibinfo{note}{ar{X}iv:quant-ph/0110034}.

\bibitem[{\citenamefont{Blume-Kohout et~al.}(2002)\citenamefont{Blume-Kohout,
  Caves, and Deutsch}}]{Blume-Kohout02}
\bibinfo{author}{\bibfnamefont{R.}~\bibnamefont{Blume-Kohout}},
  \bibinfo{author}{\bibfnamefont{C.~M.} \bibnamefont{Caves}}, \bibnamefont{and}
  \bibinfo{author}{\bibfnamefont{I.~H.} \bibnamefont{Deutsch}},
  \bibinfo{journal}{Found.\ Phys.} \textbf{\bibinfo{volume}{32}},
  \bibinfo{pages}{1641} (\bibinfo{year}{2002}),
  \bibinfo{note}{ar{X}iv:quant-ph/0204157}.

\bibitem[{\citenamefont{Nielsen and Chuang}(2000)}]{Nielsen00}
\bibinfo{author}{\bibfnamefont{M.~A.} \bibnamefont{Nielsen}} \bibnamefont{and}
  \bibinfo{author}{\bibfnamefont{I.~L.} \bibnamefont{Chuang}},
  \emph{\bibinfo{title}{Quantum computation and quantum information}}
  (\bibinfo{publisher}{Cambridge University Press},
  \bibinfo{address}{Cambridge}, \bibinfo{year}{2000}).

\bibitem[{\citenamefont{Boyer et~al.}(1998)\citenamefont{Boyer, Brassard,
  H{\o}yer, and Tapp}}]{Boyer98}
\bibinfo{author}{\bibfnamefont{M.}~\bibnamefont{Boyer}},
  \bibinfo{author}{\bibfnamefont{G.}~\bibnamefont{Brassard}},
  \bibinfo{author}{\bibfnamefont{P.}~\bibnamefont{H{\o}yer}}, \bibnamefont{and}
  \bibinfo{author}{\bibfnamefont{A.}~\bibnamefont{Tapp}},
  \bibinfo{journal}{Fortschr.\ Phys.} \textbf{\bibinfo{volume}{46}},
  \bibinfo{pages}{493} (\bibinfo{year}{1998}),
  \bibinfo{note}{ar{X}iv:quant-ph/9605034}.

\bibitem[{\citenamefont{Boykin et~al.}(2000)\citenamefont{Boykin, Mor, Pulver,
  Roychowdhury, and Vatan}}]{Boykin00}
\bibinfo{author}{\bibfnamefont{P.~O.} \bibnamefont{Boykin}},
  \bibinfo{author}{\bibfnamefont{T.}~\bibnamefont{Mor}},
  \bibinfo{author}{\bibfnamefont{M.}~\bibnamefont{Pulver}},
  \bibinfo{author}{\bibfnamefont{V.}~\bibnamefont{Roychowdhury}},
  \bibnamefont{and} \bibinfo{author}{\bibfnamefont{F.}~\bibnamefont{Vatan}},
  \bibinfo{journal}{Inform.\ Process.\ Lett.} \textbf{\bibinfo{volume}{75}},
  \bibinfo{pages}{101} (\bibinfo{year}{2000}),
  \bibinfo{note}{ar{X}iv:quant-ph/9906054}.

\bibitem[{\citenamefont{Ralph et~al.}(2002{\natexlab{a}})\citenamefont{Ralph,
  White, Munro, and Milburn}}]{Ralph01}
\bibinfo{author}{\bibfnamefont{T.~C.} \bibnamefont{Ralph}},
  \bibinfo{author}{\bibfnamefont{A.~G.} \bibnamefont{White}},
  \bibinfo{author}{\bibfnamefont{W.~J.} \bibnamefont{Munro}}, \bibnamefont{and}
  \bibinfo{author}{\bibfnamefont{G.~J.} \bibnamefont{Milburn}},
  \bibinfo{journal}{Phys.\ Rev.\ A} \textbf{\bibinfo{volume}{65}},
  \bibinfo{pages}{012314} (\bibinfo{year}{2002}{\natexlab{a}}),
  \bibinfo{note}{ar{X}iv:quant-ph/0108049}.

\bibitem[{\citenamefont{Nielsen and Chuang}(1997)}]{Nielsen97c}
\bibinfo{author}{\bibfnamefont{M.~A.} \bibnamefont{Nielsen}} \bibnamefont{and}
  \bibinfo{author}{\bibfnamefont{I.~L.} \bibnamefont{Chuang}},
  \bibinfo{journal}{Phys.\ Rev.\ Lett.} \textbf{\bibinfo{volume}{79}},
  \bibinfo{pages}{321} (\bibinfo{year}{1997}),
  \bibinfo{note}{ar{X}iv:quant-ph/9703032}.

\bibitem[{\citenamefont{Gottesman and Chuang}(1999)}]{Gottesman99}
\bibinfo{author}{\bibfnamefont{D.}~\bibnamefont{Gottesman}} \bibnamefont{and}
  \bibinfo{author}{\bibfnamefont{I.~L.} \bibnamefont{Chuang}},
  \bibinfo{journal}{Nature (London)} \textbf{\bibinfo{volume}{402}},
  \bibinfo{pages}{390} (\bibinfo{year}{1999}),
  \bibinfo{note}{ar{X}iv:quant-ph/9908010}.

\bibitem[{\citenamefont{Knill}(2002)}]{Knill02}
\bibinfo{author}{\bibfnamefont{E.}~\bibnamefont{Knill}},
  \bibinfo{journal}{Phys.\ Rev.\ A} \textbf{\bibinfo{volume}{66}},
  \bibinfo{pages}{052306} (\bibinfo{year}{2002}),
  \bibinfo{note}{ar{X}iv:quant-ph/0110144}.

\bibitem[{\citenamefont{Pittman et~al.}(2001)\citenamefont{Pittman, Jacobs, and
  Franson}}]{Pittman01}
\bibinfo{author}{\bibfnamefont{T.~B.} \bibnamefont{Pittman}},
  \bibinfo{author}{\bibfnamefont{B.~C.} \bibnamefont{Jacobs}},
  \bibnamefont{and} \bibinfo{author}{\bibfnamefont{J.~D.}
  \bibnamefont{Franson}}, \bibinfo{journal}{Phys.\ Rev.\ A}
  \textbf{\bibinfo{volume}{64}}, \bibinfo{pages}{026311}
  (\bibinfo{year}{2001}), \bibinfo{note}{ar{X}iv:quant-ph/0107091}.

\bibitem[{\citenamefont{Lund et~al.}()\citenamefont{Lund, Bell, and
  Ralph}}]{Lund03}
\bibinfo{author}{\bibfnamefont{A.~P.} \bibnamefont{Lund}},
  \bibinfo{author}{\bibfnamefont{T.~B.} \bibnamefont{Bell}}, \bibnamefont{and}
  \bibinfo{author}{\bibfnamefont{T.~C.} \bibnamefont{Ralph}},
  \bibinfo{note}{submitted to Phys.\ Rev.\ A}.

\bibitem[{\citenamefont{Hofmann and Takeuchi}(2002)}]{Hofmann02}
\bibinfo{author}{\bibfnamefont{H.~F.} \bibnamefont{Hofmann}} \bibnamefont{and}
  \bibinfo{author}{\bibfnamefont{S.}~\bibnamefont{Takeuchi}},
  \bibinfo{journal}{Phys.\ Rev.\ A} \textbf{\bibinfo{volume}{66}},
  \bibinfo{pages}{024308} (\bibinfo{year}{2002}),
  \bibinfo{note}{ar{X}iv:quant-ph/0111092}.

\bibitem[{\citenamefont{Ralph et~al.}(2002{\natexlab{b}})\citenamefont{Ralph,
  Langford, Bell, and White}}]{Ralph02}
\bibinfo{author}{\bibfnamefont{T.~C.} \bibnamefont{Ralph}},
  \bibinfo{author}{\bibfnamefont{N.~K.} \bibnamefont{Langford}},
  \bibinfo{author}{\bibfnamefont{T.~B.} \bibnamefont{Bell}}, \bibnamefont{and}
  \bibinfo{author}{\bibfnamefont{A.~G.} \bibnamefont{White}},
  \bibinfo{journal}{Phys.\ Rev.\ A} \textbf{\bibinfo{volume}{65}},
  \bibinfo{pages}{062324} (\bibinfo{year}{2002}{\natexlab{b}}),
  \bibinfo{note}{ar{X}iv:quant-ph/0112088}.

\bibitem[{\citenamefont{Biham et~al.}(2002)\citenamefont{Biham, Osborne, and
  Nielsen}}]{Biham02}
\bibinfo{author}{\bibfnamefont{O.}~\bibnamefont{Biham}},
  \bibinfo{author}{\bibfnamefont{T.~J.} \bibnamefont{Osborne}},
  \bibnamefont{and} \bibinfo{author}{\bibfnamefont{M.~A.}
  \bibnamefont{Nielsen}}, \bibinfo{journal}{Phys.\ Rev.\ A}
  \textbf{\bibinfo{volume}{65}}, \bibinfo{pages}{062312}
  (\bibinfo{year}{2002}), \bibinfo{note}{ar{X}iv:quant-ph/0112097}.

\bibitem[{\citenamefont{Fuchs}(1996)}]{Fuchs96}
\bibinfo{author}{\bibfnamefont{C.~A.} \bibnamefont{Fuchs}}, Ph.D. thesis,
  \bibinfo{school}{The University of New Mexico},
  \bibinfo{address}{Albuquerque, NM} (\bibinfo{year}{1996}),
  \bibinfo{note}{ar{X}iv:quant-ph/9601020}.

\bibitem[{\citenamefont{Wootters}(1981)}]{Wootters81}
\bibinfo{author}{\bibfnamefont{W.~K.} \bibnamefont{Wootters}},
  \bibinfo{journal}{Phys.\ Rev.\ D} \textbf{\bibinfo{volume}{23}},
  \bibinfo{pages}{357} (\bibinfo{year}{1981}).

\bibitem[{\citenamefont{Chuang and Nielsen}(1997)}]{Chuang97}
\bibinfo{author}{\bibfnamefont{I.~L.} \bibnamefont{Chuang}} \bibnamefont{and}
  \bibinfo{author}{\bibfnamefont{M.~A.} \bibnamefont{Nielsen}},
  \bibinfo{journal}{Journal of Modern Optics} \textbf{\bibinfo{volume}{44}},
  \bibinfo{pages}{2455} (\bibinfo{year}{1997}),
  \bibinfo{note}{ar{X}iv:quant-ph/9610001}.

\bibitem[{\citenamefont{Poyatos et~al.}(1997)\citenamefont{Poyatos, Cirac, and
  Zoller}}]{Poyatos97}
\bibinfo{author}{\bibfnamefont{J.~F.} \bibnamefont{Poyatos}},
  \bibinfo{author}{\bibfnamefont{J.~I.} \bibnamefont{Cirac}}, \bibnamefont{and}
  \bibinfo{author}{\bibfnamefont{P.}~\bibnamefont{Zoller}},
  \bibinfo{journal}{Phys.\ Rev.\ Lett.} \textbf{\bibinfo{volume}{78}},
  \bibinfo{pages}{390} (\bibinfo{year}{1997}),
  \bibinfo{note}{ar{X}iv:quant-ph/9611013}.

\bibitem[{\citenamefont{Knill et~al.}(2001{\natexlab{b}})\citenamefont{Knill,
  Laflamme, Martinez, and Negrevergne}}]{Knill01c}
\bibinfo{author}{\bibfnamefont{E.}~\bibnamefont{Knill}},
  \bibinfo{author}{\bibfnamefont{R.}~\bibnamefont{Laflamme}},
  \bibinfo{author}{\bibfnamefont{R.}~\bibnamefont{Martinez}}, \bibnamefont{and}
  \bibinfo{author}{\bibfnamefont{C.}~\bibnamefont{Negrevergne}},
  \bibinfo{journal}{Phys.\ Rev.\ Lett.} \textbf{\bibinfo{volume}{86}},
  \bibinfo{pages}{5811} (\bibinfo{year}{2001}{\natexlab{b}}),
  \bibinfo{note}{ar{X}iv:quant-ph/0101034}.

\end{thebibliography}

\end{document}